\documentclass[runningheads]{svmult}

\usepackage{makeidx}   
\usepackage{graphicx}  
\usepackage{subeqnar}  
\usepackage{multicol}  
\usepackage{physprbb}  
\makeindex             

\begin{document}
\title*{2dF Spectroscopy of M104 Globular Clusters}
\toctitle{2dF Spectroscopy of M104 Globular Clusters}
%
%
\titlerunning{2dF Spectroscopy of M104 Globular Clusters}
%
\author{Terry Bridges\inst{1}
\and Ken Freeman\inst{2}
\and Katherine Rhode\inst{3}
\and Steve Zepf\inst{4}}
%
\authorrunning{Bridges et al.}
%
%
\institute{Anglo-Australian Observatory
\and RSAA, Australian National University
\and Yale University
\and Michigan State University}

\maketitle              

\begin{abstract}
We present preliminary results of 2dF spectroscopy of globular clusters in
The Sombrero (M104).  We find 56 new clusters, and compile a total sample 
of 103 velocities combined with previous data.  Our 2dF data extend out to 
20 arcmin radius
($\sim$50 kpc), much further than previous studies.  In the combined sample,
we tentatively find a steep drop in the velocity dispersion with radius, 
which might possibly indicate a truncated
halo.  There is no obvious solid-body rotation over all radii, but separate
fits for those clusters inside and outside 25 kpc radius show tantalizing
evidence for counter-rotation.  The projected mass estimator with isotropic
orbits yields an M104 mass of 1.2$\times$10$^{12}$M$\odot$ inside 50 kpc, and a
(M/L)$_B$ = 30: solid evidence for a dark matter halo in this galaxy.
\end{abstract}


\section{Introduction}
Wide-field spectroscopy of globular clusters (GCs) is an excellent way 
to study galactic
mass distributions at large radii, and to learn more about the angular momentum
content of early-type galaxies (e.g. Zepf, this workshop).
GC spectra also provide cluster ages and
metallicities, which are a further means to distinguish between various models
for GC and galaxy formation.

\section{Observations}

We have used the 2dF\footnote{see www.aao.gov.au/2df} 
multi-fibre spectrograph on the AAT to obtain spectra of GC candidates in M104.
2dF, with a two degree f.o.v. and 400 fibres,
is very well-suited to wide-field studies of globular clusters in nearby galaxies.
GC candidates were obtained from KPNO Mosaic CCD BVR imaging over a 34$\times$34
arcmin field (Rhode \& Zepf 2002, in preparation).  After removal of extended
background galaxies and objects with colours outside the range of known GCs, we
have a list of 585 candidates with 19.0 $<$ V $<$ 21.5.

In April 2002, we obtained 2dF spectra for 200 of these candidates in 8 hours of 
AAT time.  We also obtained spectra for several radial velocity and Lick standard
stars.  The {\tt 2dfdr} package was used to reduce the spectra, and the IRAF/FXCOR
package used to obtain velocities via cross-correlation.  163 objects turned out to
have spectra with enough S/N to yield reliable velocities.  Of these, 56 had 
500 $<$ V $<$ 1500 km/sec, which we take to be genuine M104 GCs.  Combined with
previous WHT [1] and Keck [7] data, we
have a total sample of 103 M104 GCs.
{\bf Figure 1} shows the location of the M104 clusters from the 3 studies; it can
be seen that our 2dF data extend out to $\sim$20 arcmin radius (50 kpc for a distance
of 9 Mpc; [4]), much further out than previous studies.  

\begin{figure}[h]
\begin{center}
\includegraphics[width=.4\textwidth,angle=270]{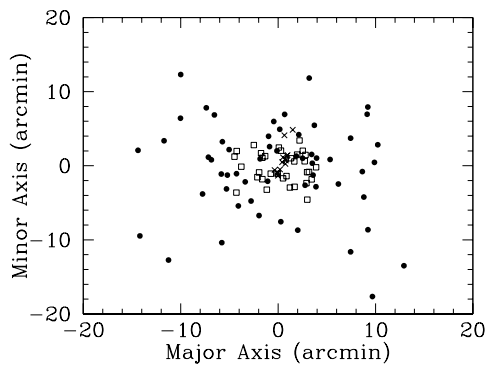}
\end{center}
\caption[]{Location of M104 GCs.  Solid points:
new 2dF GCs (April 2002); Open squares: WHT/LDSS-2 GCs [1]; 
Crosses: Keck GCs [7].  Note that the 2dF
GCs extend to a radius of $\sim$ 20 arcmin (50 kpc)}
\end{figure}

\section{Results}

\subsection{Velocity Dispersion Profile}

{\bf Figure 2a} plots velocity vs. galactocentric radius, and shows 
that the velocity dispersion decreases very quickly with radius.  To quantify this, 
we have computed smoothed velocity and velocity dispersion profiles 
with a gaussian kernel of $\sigma$ = 100 arcsec (see [9]
for more details on the technique). {\bf Figure 2b} shows that the velocity
dispersion drops from $\sim$ 225 km/sec at the
galaxy center to $\sim$ 125 km/sec at 13 arcmin (32 kpc) radius.  Such a steep drop
might indicate a truncated dark matter halo
(but see Section 3.4).

\begin{figure}[h]
\begin{minipage}[b]{0.5\textwidth}
\includegraphics[width=1.75in,angle=270]{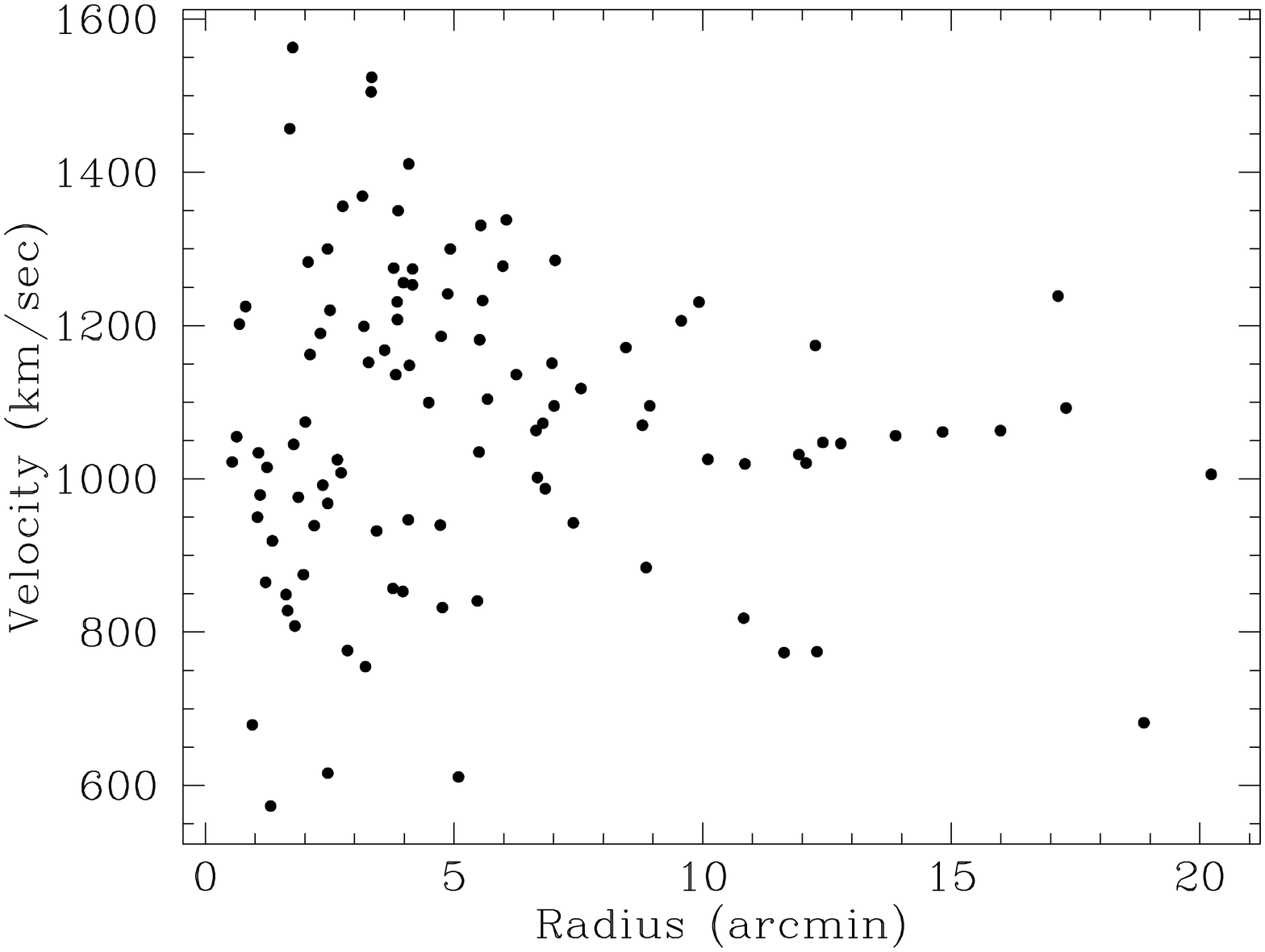}
\end{minipage}
\begin{minipage}[b]{0.5\textwidth}
\includegraphics[width=1.75in,angle=270]{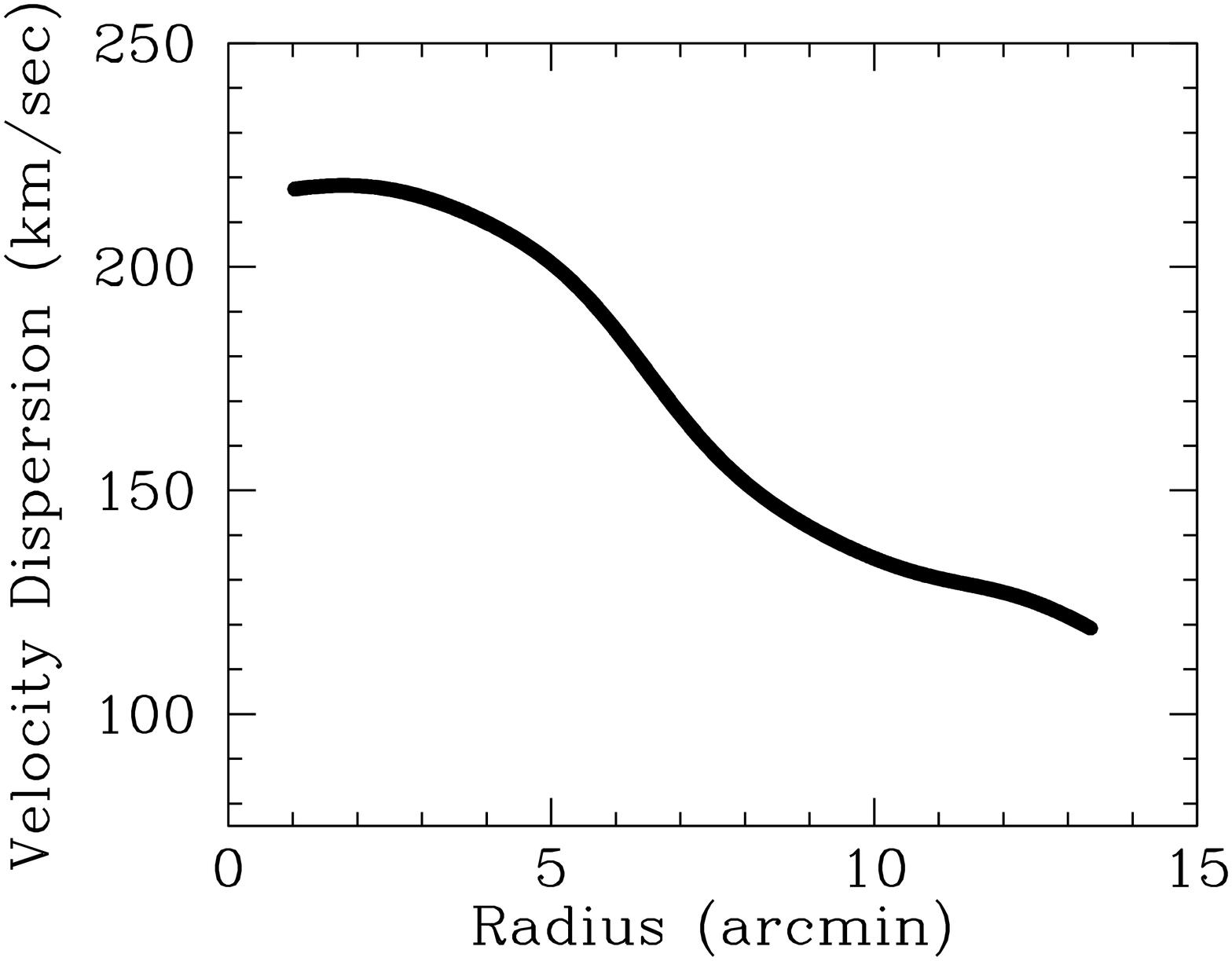}
\end{minipage}
\caption[]{{\bf (2a) (left)}: Velocity vs. galactocentric radius
for 103 confirmed M104 GCs.  {\bf (2b) (right)}: Smoothed velocity
dispersion profile, quantifying the steep drop in velocity dispersion
with radius.
}
\end{figure}

\subsection{Rotation}

In {\bf Figure 3} we plot velocity vs. azimuthal angle for the 
combined cluster sample, where $\theta$ = 0,180 corresponds to the major axis.
There is no obvious rotation (i.e. no systematic
variation of velocity with position angle).  This is confirmed by nonlinear
least squares fits for solid body rotation:

\smallskip

$V = V_{sys} + V_{rot}cos(\theta - \theta_0)$,

\smallskip

\noindent 
where V$_{sys}$ is the systematic velocity, V$_{rot}$ is the rotation amplitude,
and $\theta_0$ is the line of nodes.  The best-fit rotation is $\sim$ 60 km/sec,
but Monte Carlo simulations show that this is not significant (100 random datasets
were run through the least-squares code, with 72\% of these having a rotation
amplitude of 60 km/sec or higher).

\begin{figure}[h]
\begin{center}
\includegraphics[width=.4\textwidth,angle=270]{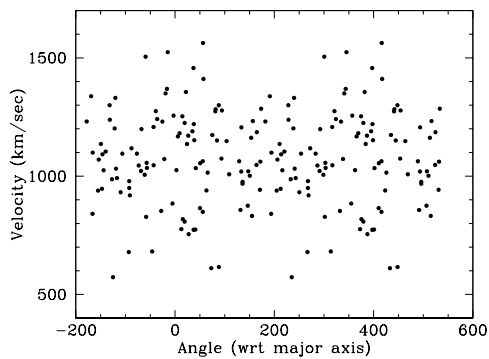}
\end{center}
\caption[]{Velocity vs azimuthal angle for M104 GCs (folded), showing
lack of rotation in combined sample.}
\end{figure}

So there appears to be no significant rotation in the combined sample over all
radii, which is in agreement with the finding of Held et al. (this workshop).
When divided by colour/metallicity, there is still no significant rotation in
either subpopulation.
However, we have investigated further by splitting up the clusters into an 
`outer' (16 GCs) and an 'inner' (87 GCs) sample, divided at a radius of 25 kpc.
{\bf Figure 4} shows that the inner and outer GCs may each have rotation,
and most interestingly, of opposite sign.  The possible rotation of the inner
clusters is in the same direction as the stellar component (e.g. [5]),
while the outer clusters would be rotating in the opposite direction.
This possible counter-rotation is extremely interesting, and may indicate a
different origin for the inner and outer GCs-- might the outer GCs have been
accreted from another galaxy?  Supporting evidence for a past interaction involving
M104 comes from the tidal features seen by [8].


\begin{figure}[h]
\begin{minipage}[b]{0.5\textwidth}
\includegraphics[width=1.75in,angle=270]{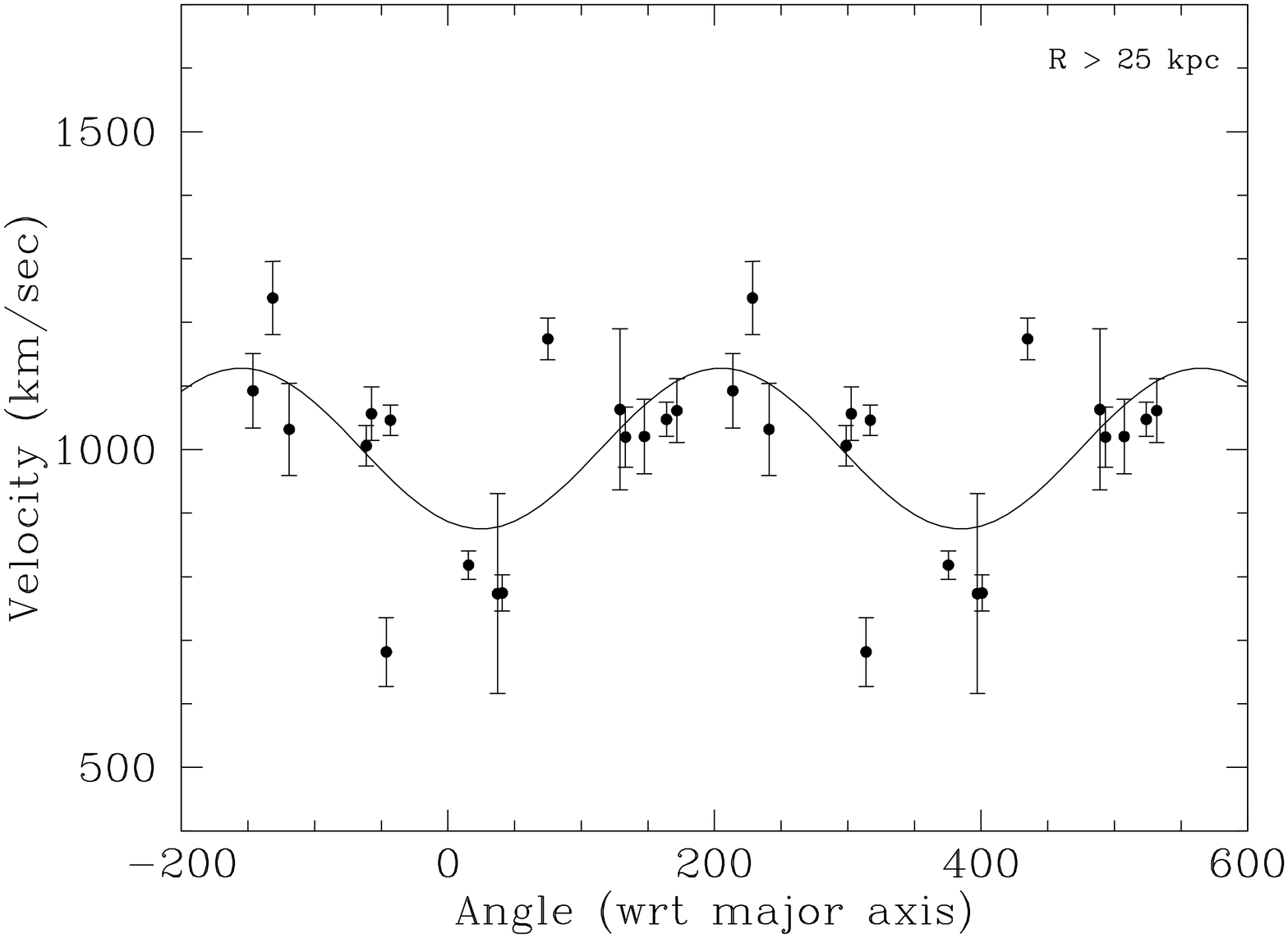}
\end{minipage}
\begin{minipage}[b]{0.5\textwidth}
\includegraphics[width=1.75in,angle=270]{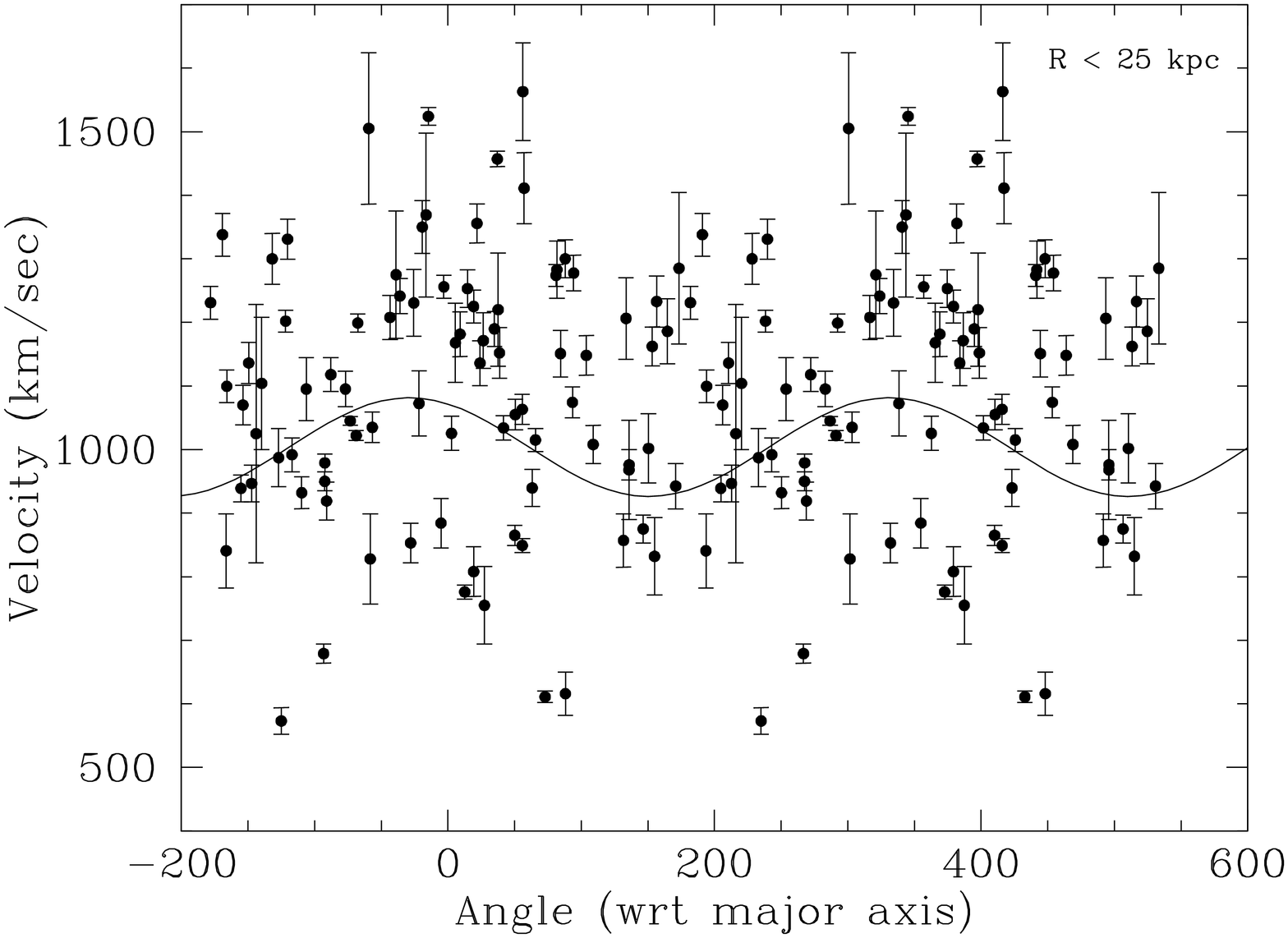}
\end{minipage}
\caption[]{Same as Figure 3, but now split by radius.  {\bf Left:}
(16) GCs with R $>$ 25 kpc, with a rotation amplitude
of 125 km/sec.  {\bf Right:}  (87) GCs with R $>$ 25 kpc, with a
rotation amplitude of 80 km/sec.  Note that the two samples are
approximately 180 degrees out of phase, as expected if they are
counter-rotating.} 
\end{figure}

\subsection{Comparison with PNe}

Freeman et al. (unpublished) have velocities for $\sim$ 250 planetary nebulae
(PNe) out to 20 kpc radius.  They find a constant velocity dispersion of 
$\sim$ 220 km/sec, and approximately constant rotation of $\sim$ 100 km/sec
out to 4 kpc; beyond 4 kpc, there is negligible rotation.  There may thus
be differences in the kinematics of the GCs and PNe, but the comparison is not
straightforward, because the two components cover different radial ranges.

\subsection{Mass of M104}

We have used the Projected Mass Estimator [6] 
to get a crude determination of the mass of M104 (to within a
factor of a few).  Assuming isotropic orbits and an extended mass distribution,
we find a mass of M= 1.2$\times$10$^{12}$ M$_\odot$ within a radius of 20 arcmin
(50 kpc).  From [2], the total integrated magnitude out to this
radius is B=8.7, and for a distance of 9 Mpc, this gives (M/L)$_B$ = 30 out
to 50 kpc.  This is solid evidence for a dark matter halo in M104.  Note that
M104 does not have extended diffuse X-ray emission, and therefore dynamical 
probes such as GCs and PNe are the only way to determine the halo mass in
this galaxy.  With larger samples of GCs, we can improve on the mass determination
by using the velocity dispersion and density profiles of the GCs and the Jeans
equation.

\section{Conclusions}

This work demonstrates the power of wide-field, multi-object spectroscopy in
the study of GC kinematics and galactic dark matter at large radius.  
It is interesting to compare our preliminary GC kinematics with those of elliptical
galaxies.  The possible rotation seen in the inner and outer M104 GCs is similar
in amplitude to that found in M49 [9] and M87 [3].  There is no evidence
for counter-rotation of GCs in M49, but the 
rotation axis of the M87 metal-poor clusters flips from the minor axis to
the major axis within 15 kpc radius ([3]).
Richtler et al (this workshop) find little rotation in the NGC 1399 GCs, except perhaps at 
large radius.  The possible counter-rotation of the outer GCs that we have found 
is very intriguing, and it may be that these GCs were accreted during a past interaction.
Further velocities are clearly required.  Hanes, Bridges, Harris,
and Gebhardt have CFHT/MOS spectra for several hundred M104 GCs, and we 
have applied for further 2dF time, so the prospects for increasing
the sample size are good.  In a related project,
Beasley, Bridges, Forbes, Harris, Harris, Mackie,
and Peng have applied for 2dF time to carry out a detailed study of the Centaurus 
A GCS.  Multi-fibre spectroscopy on 4m telescopes is quite complementary to 
programmes being carried out on 8m telescopes (e.g. the Gemini/GMOS programme
in which TJB and SEZ are involved), which naturally focus on obtaining cluster
ages and abundances over smaller fields.  However, there will be soon be many
multi-fibre spectrographs on 6-8m telescopes, with several hundred fibres
over fields of 0.5$-$1 deg (MMT/Hectospec, VLT/FLAMES, Subaru/FMOS), and the future
is looking bright indeed.

\medskip

\noindent{\bf Acknowledgements}

\medskip

TJB would like to thank the workshop organizers, in particular Markus Kissler-Patig,
Thomas Puzia, Marina Rejkuba, and Christina Stoffer, for a wonderful and very
productive workshop.  The wine and cheese session at the end was particularly
enjoyable ...

%

\end{document}